\documentclass[preprint,12pt]{elsarticle}
\usepackage{amssymb}
\usepackage{graphicx}
\usepackage{dcolumn}
\usepackage{bm}
\usepackage{amsmath}
\usepackage{latexsym}
\usepackage{epsfig}
\usepackage{amsbsy}
\usepackage{array}
\usepackage{setspace}
\usepackage{bm}
\usepackage{color}

\begin{document}

\begin{frontmatter}

\title{The spin-orbit interaction and the spin-spin interaction of photons in an inhomogeneous medium}

\author[label1]{Hehe Li}
\author[label1]{Zhigang Bu}
\author[label1]{Yuee Luo}
\author[label1]{Wenbo Chen}
\author[label1,label2]{Peiyong Ji\corref{cor1}}
\address[label1]{Department of Physics,
Shanghai University, Shanghai 200444, China}
\address[label2]{The Shanghai Key Laboratory of
Astrophysics, Shanghai 200234, China} \cortext[cor1]{Email:
$pyji@mail.shu.edu.cn$}

\begin{abstract}
By means of the optical metric, we investigate the propagation of a polarized light in an inhomogeneous medium in this paper. We find that the evolution of photons is affected by the spin-spin interaction of photons, besides the spin-orbit interaction. Due to the spin-spin interaction, there is a small deflection of the ray trajectory of the polarized light along the direction of the inhomogeneity gradient of the medium. It is different from the transverse deflection described by the spin Hall effect of photons.
\end{abstract}


\end{frontmatter}

\section{Introduction}
The optical angular momentum can be separated into three parts [1,2], the spin angular momentum, the intrinsic orbital angular momentum and the extrinsic orbital angular momentum, which are determined by the polarization, the helical phase and  the motion of center of gravity of light, respectively. The spin angular momentum can interact with the intrinsic [3] or extrinsic orbital angular momentum, and the intrinsic orbital angular momentum also can interact with the extrinsic orbital angular momentum [4,5]. One has known that there are two manifestations of the interaction between the spin angular momentum and the extrinsic orbital angular momentum of photons in an inhomogeneous medium, the Berry phase [6] and the spin Hall effect [7-25]. It is known that the inhomogeneity of the medium plays an important role in the spin-orbit interaction [7-10,17,20]. By analogy with the evolution of electrons in an external field, an inhomogeneous medium can be equivalent to a background field for the propagation of photons, the refractive index of medium $n(\textbf{r})$ and the gradient $\nabla n(\textbf{r})$ play the role of an external scalar potential and an external force affecting the motion of photons, respectively [8-10,17,20].

It is well known that a gravitational field can be equivalent to an optical medium [26,27], the optical phenomena can be approximately analysed using the theory of the geometrical optical. In 1923, Gordon pointed out that light waves experience a moving dielectric medium as a curved space which is described by Gordon optical metric [28]. By means of Gordon metric, one has investigated the propagation of light in different kinds of background medium [29-38]. This fact allows us to make an analogy between the propagation of a polarized light in an inhomogeneous medium and the evolution of spinning particles (electrons) in an external field. One has known that there is the interaction between two electrons (the spin-spin interaction) [39,40]. Is there the spin-spin interaction of photons, besides the spin-orbit interaction?

In this paper, we consider an inhomogeneous isotropic medium as a curved space by means of the optical metric. The propagation of light can be described by the Maxwell equations in curved space [27]. Based on the optical metric and the Maxwell equations in curved space, we theoretically investigate the polarization evolution of photons in the inhomogeneous isotropic medium in the geometrical optics approximation. By diagonalizing photon's Hamiltonian, we find that the photon semiclassical Hamiltonian shows a kind of the spin-spin interaction of photons, besides the spin-orbit interaction. The spin-spin interaction describes the interaction between the spin angular momenta of photons and is induced by the inhomogeneity of medium. Due to the spin-spin interaction, there is a novel and small polarization-independent deflection of the ray trajectory of the polarized light whose direction is along the inhomogeneity gradient of medium $\nabla n(\textbf{r})$. It is different from the polarization-dependent transverse deflection of the ray trajectory which originates from the spin-orbit interaction.

This paper is organized as follows. The optical metric and the Maxwell equations in curved space are introduced in Section 2. In Section 3, according to the Maxwell equations in curved space, we obtain the Hamiltonian of photons in the geometrical optics approximation using the diagonalization procedure, and find that the spin-orbit interaction and the spin-spin interaction Hamiltonian of photons correspond to the first-order and second-order correction in the geometrical optics approximation respectively. In Section 4, we investigate the influence of the spin-spin interaction on the propagation of the polarized light. We know that, because the spin-spin interaction can not distinguish the right polarization from the left polarization states, it leads to the polarization-independent correction effect. The paper is summarized in Section 5.

\section{The Maxwell equations in curved space}
\subsection{The optical effective metric}
Optical medium is experienced by electromagnetic waves as a curved space-time
which is described by an optical metric $g_{\mu\nu}$ [28]
\begin{equation}
g_{\mu\nu}=
\eta_{\mu\nu}+\left(\frac{1}{\varepsilon\mu}-1\right)u_{\mu}u_{\nu},
\end{equation}
where $\eta_{\mu\nu}=\textrm{diag}(1,-1,-1,-1)$ is the Minkowski
background metric, $u^{\mu}$ is the four-velocity of the medium with
respect to the laboratory, $\varepsilon$ and $\mu$ are the dielectric permittivity and the magnetic permittivity of the medium respectively. In the case of a static nonmagnetic isotropic medium, the optical metric can be written in the following form [41]
\begin{equation}
g_{\mu\nu}=\left(\begin{array}{cccc}\frac{1}{\varepsilon}&0&0&0\\0&-1&0&0\\0&0&-1&0\\0&0&0&-1\end{array}\right),\label{4}
\end{equation}
and the corresponding contravariant form is
\begin{equation}
g^{\mu\nu}=\left(\begin{array}{cccc}\varepsilon&0&0&0\\0&-1&0&0\\0&0&-1&0\\0&0&0&-1\end{array}\right).\label{4}
\end{equation}
From Eq. (2), one knows that the characteristics of the curved
space-time is determined by the dielectric permittivity of the medium.

\subsection{The Maxwell equations in curved space}

The evolution of photons in an inhomogeneous isotropic
medium can be described by the Maxwell equations in
curved space [27],
\begin{equation}
\frac{\partial F^{\mu\nu}}{\partial
x^{\nu}}=\frac{1}{\sqrt{-g}}\frac{\partial}{\partial
x^{\nu}}(\sqrt{-g}F^{\mu\nu})=-\frac{4\pi}{c}j^{\mu}
\end{equation}
\begin{equation}
\frac{\partial F_{\mu\nu}}{\partial x^{\lambda}}+\frac{\partial
F_{\lambda\mu}}{\partial x^{\nu}}+\frac{\partial
F_{\nu\lambda}}{\partial x^{\mu}}=0,
\end{equation}
where $F_{\mu\nu}$ is the electromagnetic field tensor, $g$ is the
determinant of the space-time metric $g_{\mu\nu}$, and $j^{\mu}$ is
the current four-vector.

The Maxwell equations (4), (5) can be
expressed in the noncovariant form
\begin{eqnarray}
\nabla\times \textbf{E}=-\frac{1}{c\sqrt{\gamma}}\frac{\partial
\left(\sqrt{\gamma}\textbf{B}\right)}{\partial t}, \\
\nabla\times
\textbf{H}=\frac{1}{c\sqrt{\gamma}}\frac{\partial\left(\sqrt{\gamma}\textbf{D}\right)}{\partial
t},\\
\nabla\cdot\textbf{B}=0,~~~~\nabla\cdot\textbf{D}=0,
\end{eqnarray}
with the constitutive equations
\begin{eqnarray}
\textbf{D}=\frac{1}{\sqrt{g_{00}}}\textbf{E}+\textbf{H}\times\textbf{g},~
\textbf{B}=\frac{1}{\sqrt{g_{00}}}\textbf{H}+\textbf{g}\times\textbf{E},
\end{eqnarray}
where $\gamma$ is the determinant of the three-dimensional metric
$\gamma_{ij}$.
In terms of Landau's definitions [27], the spatial metric is defined as
\begin{eqnarray}
\gamma^{ij}=-g^{ij},~~~
\gamma_{ij}=-g_{ij}+\frac{g_{0i}g_{0j}}{g_{00}},
\end{eqnarray}
and the covariant and contravariant components of the
three-dimensional vector $\textbf{g}$ are
\begin{eqnarray}
g_{i}=-\frac{g_{i0}}{g_{00}},~~~~
g^{i}=g^{0i}.
\end{eqnarray}

Using the definition of the wave function of photons
as ${\bf\Psi}=\textbf{D}+\mathrm{i}\textbf{B}$ [42], the
wave equation of photons is obtained from the Maxwell equations
\begin{equation}
\mathrm{i}\frac{1}{c}\frac{\partial}{\partial
t}{\bf\Psi}=\nabla\times\left(\frac{1}{\sqrt{g^{00}}}{\bf\Psi}
-\mathrm{i}\frac{1}{g^{00}}\textbf{g}\times{\bf\Psi}
\right),
\end{equation}
and the transverse condition is
\begin{equation}
\nabla\cdot{\bf{\Psi}}=0.
\end{equation}
According to the optics metric Eq. (2),
one knows that $\textbf{g}=\{g^{i}\}=0$
and $g^{00}=\varepsilon=n^{2}$, where $n$ is the refractive index of the
medium. The wave equation of photons can be represented as
\begin{equation}
\mathrm{i}\frac{1}{c}\frac{\partial}{\partial t}{\bf\Psi}
=\nabla\times\left(\frac{1}{n}{\bf\Psi}\right).
\end{equation}
The inhomogeneity of medium plays an important role in the light-matter interaction and induces the spin-orbit interaction of photons. For the circularly polarized light, the spin-orbit interaction correction is hidden in Eq. (14). In next section, we will obtain the Hamiltonian of photons from Eq. (14) by means of the iterative method and see that the spin-spin interaction also is induced by the inhomogeneity of medium.

\section{The Hamiltonian of photons}
In this section, we consider a monochromatic polarized light propagates in a stationary inhomogeneous isotropic medium. Assuming that the inhomogeneity of medium is smooth, a parameter of geometrical optics
$\chi=\lambda/L\ll 1$ can be introduced, where $\lambda=\lambda_{0}/n$ characterizes the wavelength of light in medium, $\lambda_{0}=c/\omega$ is the wavelength in vacuum, $\omega$ is the angular frequency of light, and $L=n/|\nabla n|$ is the characteristic scale of inhomogeneous medium. In the following, we will investigate the propagation of photons in the geometrical optics approximation.

Considering a monochromatic light, one obtains $\partial\mathbf{\Psi}/\partial t=-i\omega\mathbf{\Psi}$. By virtue of the iterative method, we obtain the following equation from Eq. (14),
\begin{align}
{\frac{\omega^{2}}{c^{2}}{\bf{\Psi}}}&=
\frac{1}{n^{2}}\nabla\times(\nabla\times{\bf\Psi})
+\frac{2}{n}\nabla\frac{1}{n}\times(\nabla\times{\bf\Psi})\nonumber
\\
&\quad+\frac{1}{n}\nabla\times\left(\nabla\frac{1}{n}\times{\bf\Psi}\right)
+\nabla\frac{1}{n}\times\left(\nabla\frac{1}{n}\times{\bf\Psi}\right).
\end{align}

Introducing the spin operator $\textbf{s}=(s^{i})_{jk}=-\mathrm{i}\epsilon^{ijk}$ and the momentum operator $\textbf{p}=-\mathrm{i}\lambda_{0}\nabla$ of photons, a relativistic Schr$\ddot{\mathrm{o}}$dinger-type equation of photons is obtained as follows,
\begin{eqnarray}
\left\{\textbf{p}^{2}-n^{2}-\lambda_{0}\textbf{s}\cdot\frac{\nabla n\times\textbf{p}}{n}
-\lambda_{0}^{2}\frac{(\textbf{s}\cdot\nabla n)(\textbf{s}\cdot\nabla n)}{n^{2}}\right\}{\bf{\Psi}}
=0.
\end{eqnarray}
The transverse condition (13) has been used and the imaginary terms have been ignored in Eq. (16). The Hamiltonian of photons can be written as the following form
\begin{equation}
\hat{H}=\textbf{p}^{2}-n^{2}-\lambda_{0}\textbf{s}\cdot\frac{\nabla n\times\textbf{p}}{n}
-\lambda_{0}^{2}\frac{(\textbf{s}\cdot\nabla n)(\textbf{s}\cdot\nabla n)}{n^{2}}.
\end{equation}
Obviously, the Hamiltonian (17) is in the nondiagonal form and mixes up different polarization eigenstates of photons. To find eigenstates of photons and to describe their evolution, we diagonalize the Hamiltonian (17) using the transformation matrix $U(p)$ which is built of the orthogonal set of unit eigenvectors derived from $\textbf{s}\cdot\textbf{p}|{\bf\Psi}\rangle=m p|{\bf\Psi}\rangle$, where $m=\pm1, 0$ correspond to the transverse states and the longitudinal state of electromagnetic wave respectively,
\begin{equation}
U(p)=\left(\begin{array}{ccc}\frac{\mathrm{i}p_{x}p_{z}+p_{y}p}{\sqrt{2}p\sqrt{p_{x}^{2}+p_{y}^{2}}}&
\frac{\mathrm{i}p_{x}p_{z}-p_{y}p}{\sqrt{2}p\sqrt{p_{x}^{2}+p_{y}^{2}}}&
\frac{p_{x}}{p}\\\frac{\mathrm{i}p_{y}p_{z}-p_{x}p}{\sqrt{2}p\sqrt{p_{x}^{2}+p_{y}^{2}}}&
\frac{\mathrm{i}p_{y}p_{z}+p_{x}p}{\sqrt{2}p\sqrt{p_{x}^{2}+p_{y}^{2}}}&\frac{p_{y}}{p}\\
\frac{-\mathrm{i}\sqrt{p_{x}^{2}+p_{y}^{2}}}{\sqrt{2}p}&
\frac{-\mathrm{i}\sqrt{p_{x}^{2}+p_{y}^{2}}}{\sqrt{2}p}&\frac{p_{z}}{p}\end{array}\right).\label{4}
\end{equation}
Using $U(p)^{\dag}\hat{H}U(p)=H_{ij}$, where $i=1,2,3$, the Hamiltonian of photons corresponding to the right and left polarization states can be obtained, $H^{+}=H_{11}$, $H^{-}=H_{22}$,
\begin{equation}
H^{\sigma}=\textbf{p}^{2}-n^{2}-\mathbf{\Sigma}\cdot\frac{\nabla n
\times\textbf{p}}{n^{3}}-\lambda_{0}^{2}\frac{1}{2n^{4}}\left[(\nabla n\cdot\nabla n)+\lambda_{0}^{-2}(\mathbf{\Sigma}\cdot\nabla
n)(\mathbf{\Sigma}\cdot\nabla n)\right],
\end{equation}
where $\mathbf{\Sigma}=\sigma\lambda_{0}\textbf{p}/p$ is the spin angular momentum of light, and $\sigma=\pm1$ denotes the wave helicity indicating the two spin states of photons. We can see that the first three terms in Eq. (19) are the zero-order and first-order approximation in $\chi$, which have been obtained in [7-10,12-14,17,20]. One has known that the first-order correction in $\chi$ in the Hamiltonian Eq. (19) describes the spin-orbit interaction of photons in an inhomogeneous medium. The fourth term in Eq. (19) is a novel term and a second-order correction in $\chi$ and describes the interaction between the spin angular momentum of photons. By analogy with the spin-spin interaction of two electrons [39,40], the fourth term in Eq. (19) can be called the spin-spin interaction Hamiltonian.

In the smoothly inhomogeneous medium, electromagnetic wave is almost transverse wave and the longitudinal component of electromagnetic wave is absent. Therefore, the components of the Hamiltonian $H_{33}$, which corresponds to the longitudinal state, should be neglected. The cross-components, $H_{12}$ and $H_{21}$, describe the transitions between the right and left polarization states [43,44], while $H_{13}$, $H_{23}$, $H_{31}$ and $H_{32}$, describe the transitions between transverse and longitudinal wave states. The cross-components $H_{ij} (i\neq j)$ is proportional to $\chi^{2}$ and neglected in [10,14].

\section{The influence of the spin-spin interaction on the propagation of photons}
One has known that, the spin Hall effect and the Berry phase are two manifestations of the spin-orbit interaction of photons, which displays a splitting of the trajectories of different circularly polarized lights and a rotation of the polarization ellipse respectively. In this section, we will investigate the influence of the spin-spin interaction on the ray trajectory of the polarized light.

\subsection{The influence of the spin-spin interaction on the ray trajectory of the circularly polarized light}

Using the Hamiltonian (19) and the Canonical equations, the motion equation
of photons can be obtained
\begin{equation}
\dot{\textbf{r}}=\frac{2\textbf{p}}{n}-\sigma\lambda_{0}\frac{\nabla
n\times\textbf{p}}{n^{2}p}-\lambda_{0}^{2}\frac{\nabla
n\cdot\textbf{p}}{n^{3}p^{2}}\nabla n,
\end{equation}
where dot denotes the derivative with respect to the ray parameter $s$, which is connected with the ray length, $dl=nds$. The second term in Eq. (20) originates from the Hamiltonian of the spin-orbit interaction Hamiltonian of photons and describes a polarization-dependent transverse deflection of the ray trajectory which is responsible for the spin Hall effect of photons. The third term originates from the spin-spin interaction Hamiltonian of photons and describes a polarization-independent deflection of the ray trajectory in the direction of $\nabla
n$ (see Fig. 1). This is a novel and important correction, and it is completely different from the transverse deflection in the direction of $\nabla
n\times\textbf{p}$ (the second term in Eq. (20)).

It is worthy to be noticed that the deflection induced by the spin-spin interaction is proportional to the second-order geometrical optics parameter $\chi$,it is a tiny effect and ignored in the previous works [7,8,12,14,17,20].

\subsection{The influence of the spin-spin interaction on the phase evolution of the circularly polarized light}

The Berry phase is a nonintegrable phase, which arises from the adiabatic transport of a system around a closed path in parameter space, and written as the following form according to Berry's definition [6],
\begin{equation}
\gamma_{m}(C)=\oint_{C}d\emph{\textbf{R}}\cdot\textbf{A}_{m}(\textbf{R}),
\end{equation}
where $\textbf{A}_{m}(\textbf{R})=\mathrm{i}\langle{m,\emph{\textbf{R}}}|\nabla_{\emph{\textbf{R}}}|{m,\emph{\textbf{R}}}\rangle$ is the Berry connection, $|{m,\emph{\textbf{R}}}\rangle$ is the eigenstates in the parameter $\textbf{R}$ space. One knows that the photon's spin state $|{\sigma,\emph{\textbf{p}}}\rangle$ satisfies the function $\textbf{s}\cdot\textbf{p}|{\sigma,\emph{\textbf{p}}}\rangle=\sigma p|{\sigma,\emph{\textbf{p}}}\rangle$, the photon's spin state $|{\sigma,\emph{\textbf{p}}}\rangle$ are obtained
\begin{equation}
|{+,\emph{\textbf{p}}}\rangle=\frac{(-\mathrm{i})}{\sqrt{2}p\sqrt{p_{x}^{2}+p_{y}^{2}}}\left(\begin{array}{ccc}-p_{x}p_{z}+\mathrm{i}p_{y}p
\\-p_{y}p_{z}-\mathrm{i}p_{x}p\\p_{x}^{2}+p_{y}^{2}\end{array}\right),\label{3}
\end{equation}
\begin{equation}
|{-,\emph{\textbf{p}}}\rangle=\frac{(-\mathrm{i})}{\sqrt{2}
p\sqrt{p_{x}^{2}+p_{y}^{2}}}\left(\begin{array}{ccc}-p_{x}p_{z}-\mathrm{i}p_{y}
p\\-p_{y}p_{z}+\mathrm{i}p_{x}p\\p_{x}^{2}+p_{y}^{2}\end{array}\right).\label{3}
\end{equation}
The Berry connection $\emph{\textbf{A}}$ (the effective vector-potential) in the momentum $\textbf{p}$ space is obtained as following [10],
\begin{equation}
\emph{\textbf{A}}_{\pm}=\pm(-\frac{p_{y}p_{z}}{p(p_{x}^{2}+p_{y}^{2})},\frac{p_{x}p_{z}}{p(p_{x}^{2}+p_{y}^{2})},0),
\end{equation}
the corresponding Berry curvatures (the field tensors) can be calculated directly,
\begin{equation}
\emph{\textbf{V}}_{\pm}=\nabla_{\textbf{p}}\times\textbf{A}_{\pm}=\mp\frac{\emph{\textbf{p}}}{p^{3}}.
\end{equation}
Eq. (25) presents a magnetic-monopole-like field in the momentum $\textbf{p}$ space [10,15,17,20]. The Berry phase can be calculated as
\begin{equation}
\gamma_{\sigma}(C)=\oint_{C}d\emph{\textbf{p}}\cdot\textbf{A}_{\pm}(\textbf{p})=\int_{S}d\emph{\textbf{S}}\cdot\textbf{V}_{\pm}(\textbf{p}).
\end{equation}
Here $S$ is a surface spanned over the closed contour $C$. The Berry phase is equal to the contour integral of the effective vector-potential $\emph{\textbf{A}}$ or to the flux of the effective field $\emph{\textbf{V}}$ through this contour. It displays the rotation of the polarization ellipse and has been measured experimentally in a single-mode helical optical fibres [45] and for helical rays in a multimode rectilinear waveguide [17].

The Berry phase is a manifestations of the spin-orbit interaction of photons. The transverse wave
state represents a polarization degeneration of electromagnetic waves and corresponds to the spin degeneration of energy levels of photons. Owing to the spin-orbit interaction, the different independent eigenstates of photons can be distinguished in the momentum $\textbf{p}$ space. On the contrary, the spin-spin interaction of photons can not distinguish the right polarization from the left polarization states, because it is independent of the polarizations of photons. The spin-spin interaction just makes a small phase contribution $\phi_{SS}$ to the dynamical phase,
\begin{equation}
\phi_{SS}=\lambda_{0}\int\frac{1}{n^{4}}\left[(\nabla n)^{2}-\frac{(\nabla n\cdot\textbf{p})^{2}}{p^{2}}\right]dt.
\end{equation}

\section{Conclusions}
In this paper, we investigate the propagation of the polarized light (photons) in an inhomogeneous isotropic medium based on the theory of optical metric and the Maxwell equations in curved space. We find that the propagation of the polarized light is affected by the spin-spin interaction, besides the spin-orbit interaction. From the Hamiltonian (19), one knows that the inhomogeneity of medium plays an important role in the spin-orbit interaction and the spin-spin interaction of photons. The spin-orbit interaction leads to a polarization-dependent transverse deflection of the ray trajectory of the circularly polarized light, the direction of deflection is orthogonal both to the photon's momentum $\textbf{p}$ and inhomogeneity gradient $\nabla
n$. The spin-spin interaction leads to a small and polarization-independent deflection of the ray trajectory in the direction of inhomogeneity gradient $\nabla
n$. This is slightly different from the transverse deflection of the ray trajectory of the circularly polarized light which is responsible for the spin Hall effect of photons. It also should be noted that the spin-spin interaction is independent of the polarization of photons and induces a polarization-independent correction phase to the dynamical phase of photons, it doesn't affect the rotation of the polarization ellipse (the Berry phase).

The polarization-dependent transverse displacement due to the spin Hall effect of photons has been observed in the experiment [17,18].  According to the motion equation of photons (20), the deflection of the ray trajectory, which originates from the spin-spin interaction, is the second-order correction in the geometrical optics approximation and is a very tiny optical phenomena. Even so, the spin-spin interaction induced by the inhomogeneity of medium further confirms the nature of wave-particle duality of light.
\\\\
\textbf{Acknowledgments}
\\
This work was partly supported by the Shanghai Leading
Academic Discipline Project (Project No. S30105) and the Shanghai
Research Foundation (Grant No. 07dz22020).

\newpage
\section*{References}
\label{}

\newpage

\begin{figure*}[htb]
\includegraphics[scale=.6]{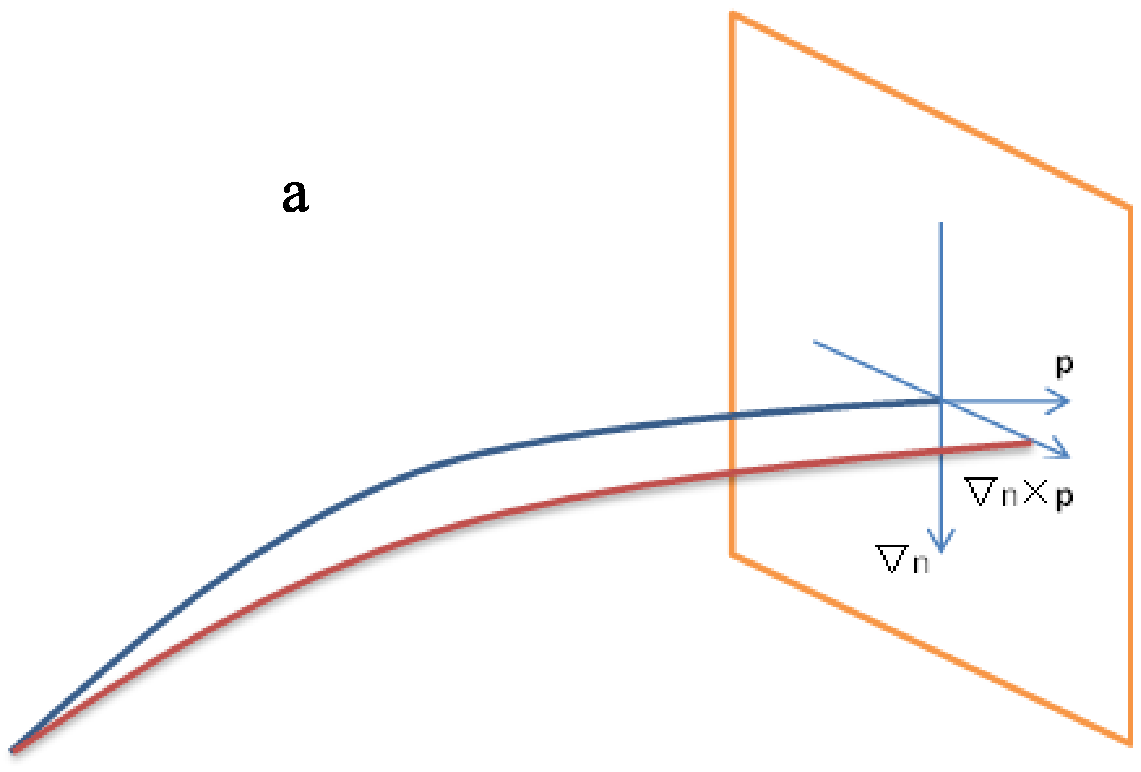}
\includegraphics[scale=.6]{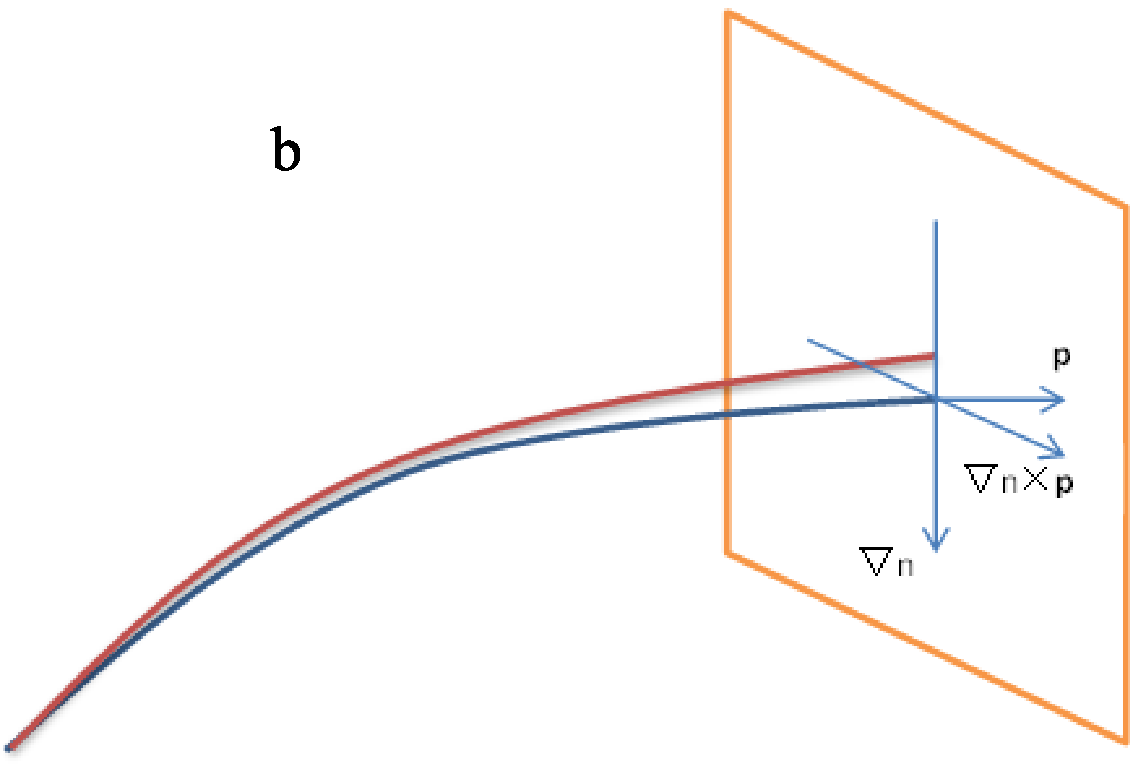}
\caption{\label{fig:epsart} Schematic picture of the deflection of the ray trajectory of the polarized light. The blue line depicts the ray trajectory in the zero-order approximation in $\chi$, the red line depicts the ray trajectory with the correction of spin interaction. \textbf{a}, The deflection of the ray trajectory originated from the spin-orbit interaction is along the direction of $\nabla n\times\textbf{p}$ and called the spin Hall effect. \textbf{b}, The deflection of the ray trajectory originated from the spin-spin interaction is along the direction of $\nabla n$.}
\end{figure*}

\end{document}